\documentclass[letterpaper,twocolumn,10pt]{article}
\usepackage{usenix2019_v3}

\usepackage{tikz}
\usepackage{amsmath}
\usepackage{cleveref}
\usepackage{listings}
\usepackage{subcaption}
\usepackage{microtype}
\usepackage{amssymb}
\usepackage{enumitem}
\usepackage{algorithm}
\usepackage{hyperref}
\usepackage{algpseudocodex}
\usepackage{xurl}

\usepackage[inkscapelatex=false]{svg}

\usepackage{duckuments}

\usepackage{mathtools}

\newcommand{\Ilp}{$\mathbb{Z}$LP}
\newcommand{\loverii}{\lceil L/I \rceil}

\usepackage{filecontents}

\definecolor{todocolor}{rgb}{0.8,0,0}
\definecolor{editcolor}{rgb}{0,0,0.8}

\newcommand{\IGNORE}[1]{}

\definecolor{keywordcolor}{rgb}{0.5,0,0.5}
\definecolor{textgray}{gray}{0.4}
\definecolor{mygray}{rgb}{0.5,0.5,0.5}
\hypersetup{
    breaklinks=true,
    colorlinks=true,
    urlcolor=blue,
}

\newcommand{\name}{Twill}

\begin{document}

\date{}

\title{\Large \bf {Optimal Software Pipelining and Warp Specialization for Tensor Core GPUs} }

\author{
{\rm Rupanshu Soi\thanks{Equal contribution.}\, \thanks{This work was done at NVIDIA.}}\\ 
Stanford University
\and
{\rm Rohan Yadav\footnotemark[1]\, \footnotemark[2]}\\
Stanford University
\and
{\rm Fredrik Kjolstad}\\
Stanford University
\and
{\rm Alex Aiken\footnotemark[2]}\\
Stanford University
\and
{\rm Maryam Mehri Dehnavi}\\
NVIDIA
\and
{\rm Michael Garland}\\
NVIDIA
\and
{\rm Michael Bauer}\\
NVIDIA
} 

\maketitle

\begin{abstract}

GPU architectures have continued to grow in complexity, with recent incarnations introducing increasingly powerful fixed-function units for matrix multiplication and data movement to accompany highly parallel general-purpose cores.
To fully leverage these machines, software must use sophisticated schedules that maximally utilize all hardware resources.
Since realizing such schedules is complex, both programmers and compilers routinely employ program transformations, such as software pipelining (SWP) and warp specialization (WS), to do so in practice.
%
However, determining how best to use SWP and WS in combination is a challenging problem that is currently handled through a mix of brittle compilation heuristics and fallible human intuition, with little insight into the space of solutions.
%
%
%
%
%
To remedy this situation, we introduce a novel formulation of SWP and WS as a joint optimization problem that can be solved holistically by off-the-shelf constraint solvers.
We reify our approach in \name{}, the first system that automatically derives optimal
SWP and WS schedules for a large class of iterative programs. 
%
%
\name{} is heuristic-free, easily extensible to new GPU architectures,
and guaranteed to produce optimal schedules.
We show that \name{} can rediscover, and thereby prove optimal,
the SWP and WS schedules manually developed by 
experts for Flash Attention on both the NVIDIA Hopper and Blackwell 
GPU architectures.

\end{abstract}

\section{Introduction}

Driven by the insatiable appetite of machine learning for performance, recent
GPUs continue to expand the scale 
and capabilities
of fixed-function units for
matrix multiplication (GEMM), called Tensor Cores (TCs) on NVIDIA GPUs, and bulk data
movement.  
%
%
As these fixed-function units become more powerful, the relative performance ratios of data movement and FLOPS between general-purpose and fixed-function matrix units change dramatically, often by multiplicative factors.
Additionally, the interfaces for targeting fixed-function units routinely undergo significant alterations across generations, ranging from modifications to where data can be placed, to how many threads are required to issue operations, and even to the model of asynchronous execution.
Consequently, every program targeting TC GPUs may have a different optimal \emph{schedule} for each architecture generation to fully leverage the available general-purpose and fixed-function units~\cite{flash-attention-3, cutlass-b100-fmha, flash-attention-4}.
%
%
%

A common solution to this performance portability problem for iterative programs is \emph{software pipelining} (SWP)~\cite{aiken-pipelining, lam-modulo-scheduling,dragon-book}.
SWP permits programmers to write a simple loop to describe a computation, which a compiler can then automatically transform to exploit instruction-level parallelism within and across iterations to derive a schedule that maximizes utilization of all the functional units.
Importantly, when compiling the same loop for different machines, the compiler will tailor a custom schedule based on the underlying constraints for each architecture to ensure that the same program runs efficiently in all settings~\cite{dragon-book}.  
To realize a particular SWP schedule, the compiler must synthesize a new loop, but the requirements imposed by new TC GPUs, such as the need for many threads to cooperatively issue TC operations~\cite{hopper-white-paper,blackwell-blogpost}, 
preclude the use of a standard sequential loop to express most schedules.

To realize SWP schedules for TC GPUs, both compilers~\cite{triton, tawa, cypress, tlx, pallas, tile-lang} and developers~\cite{cutlass, thunderkittens} have adopted \emph{warp specialization} (WS)~\cite{cuda-dma,singe} as a programming paradigm.
%
Instead of the standard data-parallel style of GPU programming that assigns the same computation to each thread, in WS, subsets of threads called \emph{warps} collaboratively execute different parts 
%
of a program, exchanging data through common memories and synchronizing where necessary.
%
While WS is a necessary program transformation for achieving SWP schedules for TC GPUs, it comes with performance trade-offs (e.g., extra communication and synchronization) that obscure how best to deploy it in practice.


Currently, approaches for both SWP and WS on TC GPUs are 
derived either by human intuition~\cite{cutlass, flash-attention-3, flash-attention-4}
or through compiler heuristics~\cite{triton, cypress, tawa}.
Moreover, the interaction between SWP and WS is 
poorly understood, with no technical framework for reasoning about the optimality
of combined solutions.
%
%
%
%
The hazards of this situation can be observed in the case of
Flash Attention~\cite{flash-attention, flash-attention-2}, where a year transpired between the release of
Hopper and the development of Flash Attention 3~\cite{flash-attention-3},
which proposed a custom SWP schedule and WS strategy for Hopper, wasting
countless compute cycles on critical inference workloads in the interim.
%
%
%
%
%

To rectify this situation, we show that the problems of determining the best SWP schedule and WS strategy can and should be solved simultaneously.  
We observe that determining
a SWP schedule to maximize resource utilization 
can be reduced to a traditional \emph{modulo scheduling}
problem~\cite{lam-modulo-scheduling, rau-modulo-scheduling}, that
can be solved \emph{optimally}~\cite{gao-ilp-scheduling, stoutchinin-ilp}, 
yielding the maximum possible throughput 
for any given machine model. 
%
We then explain why WS is a dependent program transformation that must be jointly considered to compute realizable SWP schedules for recent TC GPU architectures. 
%
%
We extend the approach of modulo scheduling to formulate the
problem of determining an optimal SWP schedule and WS strategy as a unified constraint satisfaction
problem that can be solved holistically by off-the-shelf Satisfiability
Modulo Theories (SMT) solvers~\cite{smt-solvers}.

To demonstrate our approach, we introduce \name{}\footnote{Twill is a textile weave that interlaces warp and weft threads in a distinctive diagonal pattern.}, a system that automatically
discovers optimal SWP schedules and WS strategies from high-level, tile-based descriptions of
loops with simple control flow.
\name{} generates optimal schedules for different GPU architectures simply 
by altering machine-specific constraints corresponding to the easily quantifiable
costs of various GPU operations. 
In contrast to all existing systems that perform SWP and WS of which we are
aware~\cite{triton, cypress, tawa, tile-lang}, our implementation is heuristic-free,
easily extensible to new GPU architectures,
and guaranteed to yield an optimal schedule for a large and important class of iterative programs (i.e., singly-nested loops without additional control flow).
%
The specific contributions of this work are:
\begin{itemize}[nosep]
    \item A mapping of the problem of constructing a SWP schedule for TC GPUs to modulo scheduling.
    \item A constraint-based formulation of SWP and WS as a joint optimization problem.
    \item \name{}, a system that implements our approach and discovers optimal SWP schedules and WS strategies for real programs.
\end{itemize}

We evaluate \name{} by applying it to the Flash Attention algorithm and derive schedules for
both the NVIDIA Hopper and Blackwell architectures.
We show that \name{} can automatically generate schedules that match the proposed algorithms
in Flash Attention 3~\cite{flash-attention-3} and Flash Attention 4~\cite{flash-attention-4}
from a high-level description of attention.
We then implement these schedules and show that the resulting programs
can come within 1\% of the performance of hand-tuned implementations from libraries like cuDNN and Flash Attention.

\section{Background on GPU Architecture}\label{sec:background}


NVIDIA GPUs consist of a number of independent processors called streaming multiprocessors (SMs).
%
%
%
Each SM contains independent functional units to support general-purpose floating-point or integer arithmetic, as well as load and store units for memory operations.
SMs also contain several kinds of memory including a register file for storing thread-local data,  a software-managed scratchpad called \emph{shared memory} that can be accessed by all threads on an SM, and an off-chip \emph{global memory} that can be accessed from any SM.

An SM executes groups of 32 threads called \emph{warps}.
Warps execute in a \emph{single instruction, multiple threads} (SIMT)
model: every thread in a warp has its own instruction stream, but only a subset of threads that agree on a common instruction to execute will issue in a given cycle, while the remaining threads are masked off and prevented from executing. 
Each thread executes \emph{in-order}: if a thread's next instruction is blocked on
an unresolved dependency or synchronization, it cannot issue any further instructions.
Hopper and Blackwell SMs have four \emph{execution contexts} that can host active warps, so
at most four warps can issue instructions each cycle.
When there are more warps than execution contexts, the SM's \emph{warp scheduler} dynamically selects up to four ready warps from which to issue instructions.

Modern GPUs augment this architecture by incorporating asynchronous accelerators for operating on entire tiles of data.
Tensor Cores (TCs) for accelerating the performance of GEMM operations were first introduced in the Volta architecture~\cite{volta-white-paper}.
More recently, the Hopper architecture~\cite{hopper-white-paper} introduced the Tensor Memory Accelerator (TMA) unit for asynchronously moving tiles of data between global and shared memory.
%
Since they operate over large tiles, a single TMA or TC instruction may execute for thousands of clock cycles, in stark contrast to most floating-point or integer arithmetic instructions which often only execute for tens of cycles.

The Blackwell architecture~\cite{blackwell-blogpost} is similar to Hopper, but with two important enhancements.
The Blackwell TC supports larger tile sizes with higher throughput, and each SM contains a new kind of memory called Tensor Memory where inputs to the TC may be sourced and accumulators stored.
Performing general computations on Blackwell TC accumulators requires explicit data movement from the Tensor Memory into the register file.
While conceptually straightforward, these modifications induce large perturbations in how programs must be structured for peak performance on Blackwell.

\section{Scheduling for Tensor Core GPUs}
\label{sec:scheduling}

\begin{figure*}[t]

\begin{subfigure}[b]{0.235\textwidth}
\centering
\begin{lstlisting}
O = zeros()
for i:
  S = gemm(Q, K[i])
  P = exp(S)
  O += gemm(P, V[i])
\end{lstlisting}
\caption{Simplified Flash Attention pseudocode for 1 SM.}
\label{fig:simpl-attn}
\end{subfigure}\hfill%
\begin{subfigure}[b]{0.235\textwidth}
\centering
    \includegraphics[width=\textwidth]{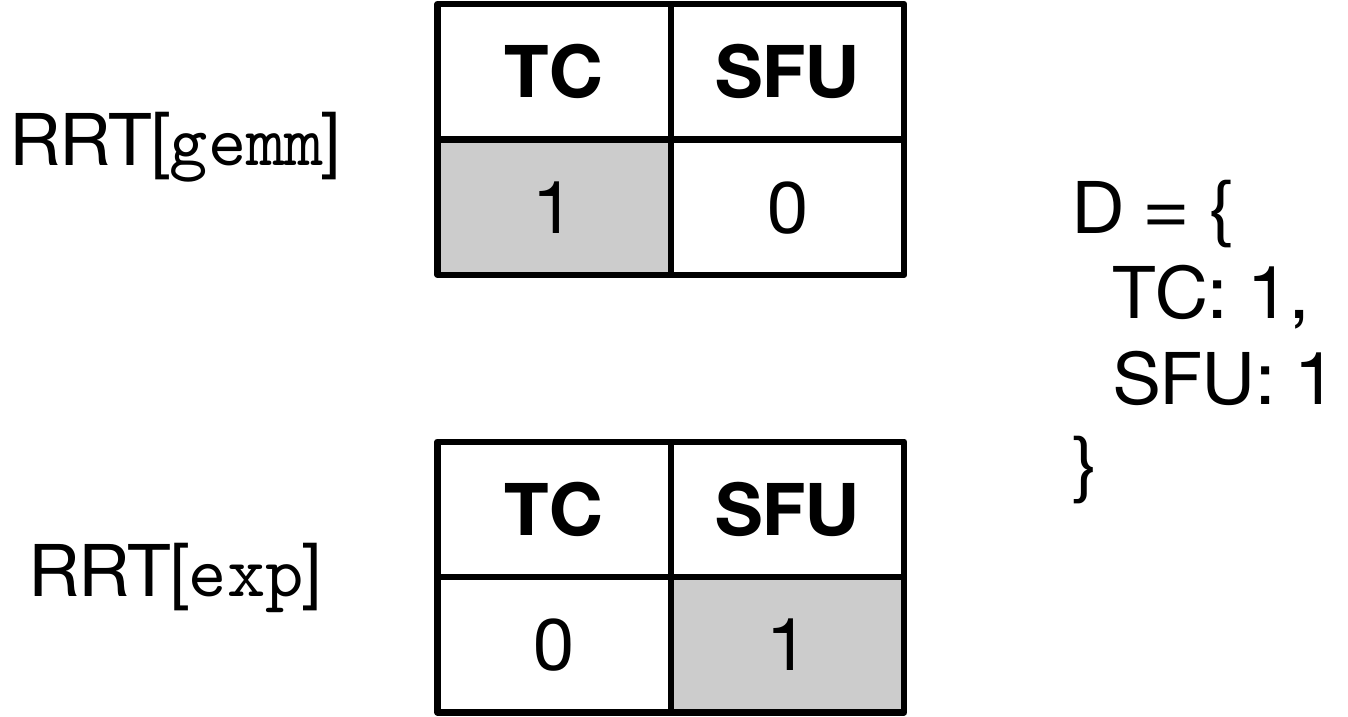}
\caption{Instruction RRTs and Hopper machine description.}
\label{fig:attn-rrts}
\end{subfigure}\hfill%
\begin{subfigure}[b]{0.235\textwidth}
\centering
\includegraphics[width=\textwidth]{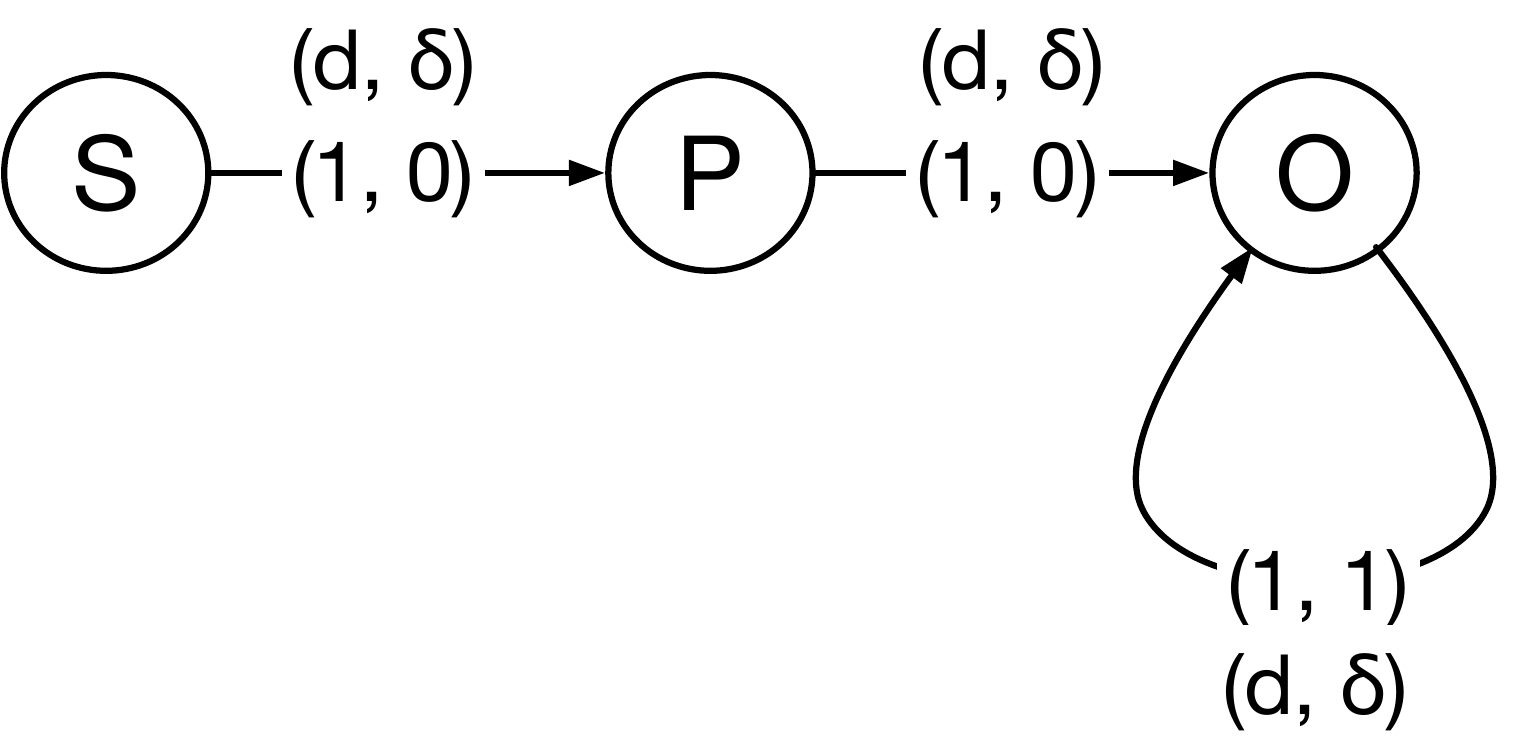}
\caption{Loop dependence graph representation of \Cref{fig:simpl-attn}.}
\label{fig:attn-depgraph}
\end{subfigure}\hfill%
\begin{subfigure}[b]{0.235\textwidth}
\centering
    \includegraphics[width=\textwidth]{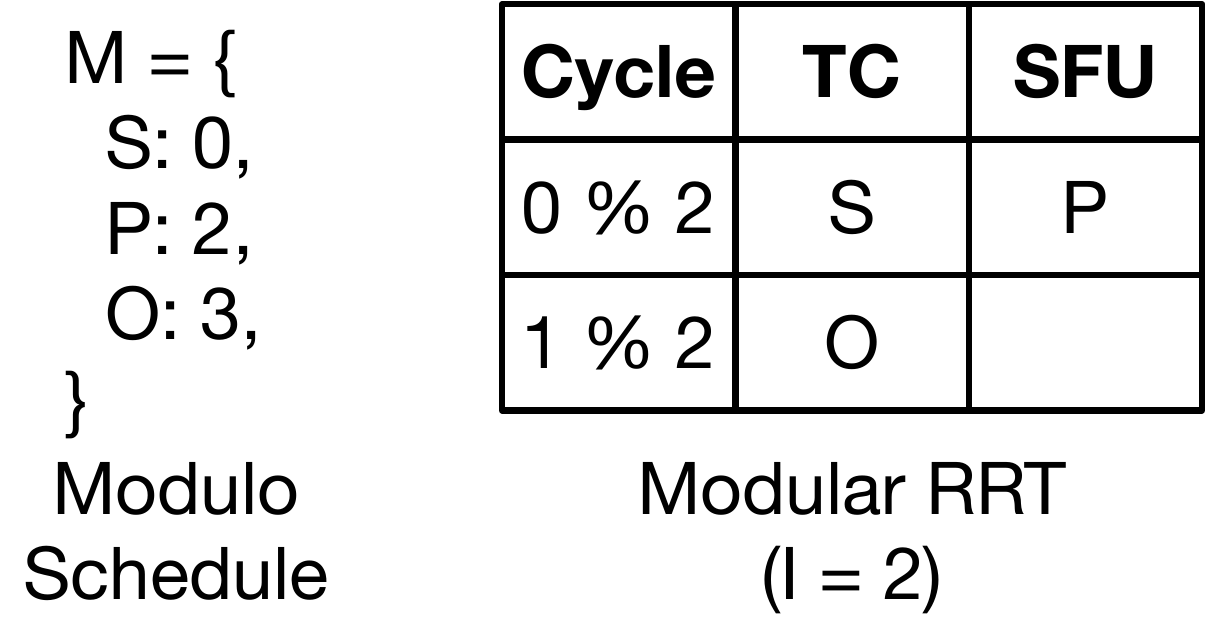}
    \caption{Valid modulo schedule and modular RRT with $I=2$, $L=4$.}
\label{fig:attn-modulo-schedule}
\end{subfigure}

\quad

\begin{subfigure}[b]{0.235\textwidth}
\centering
    \includegraphics[width=0.65\textwidth]{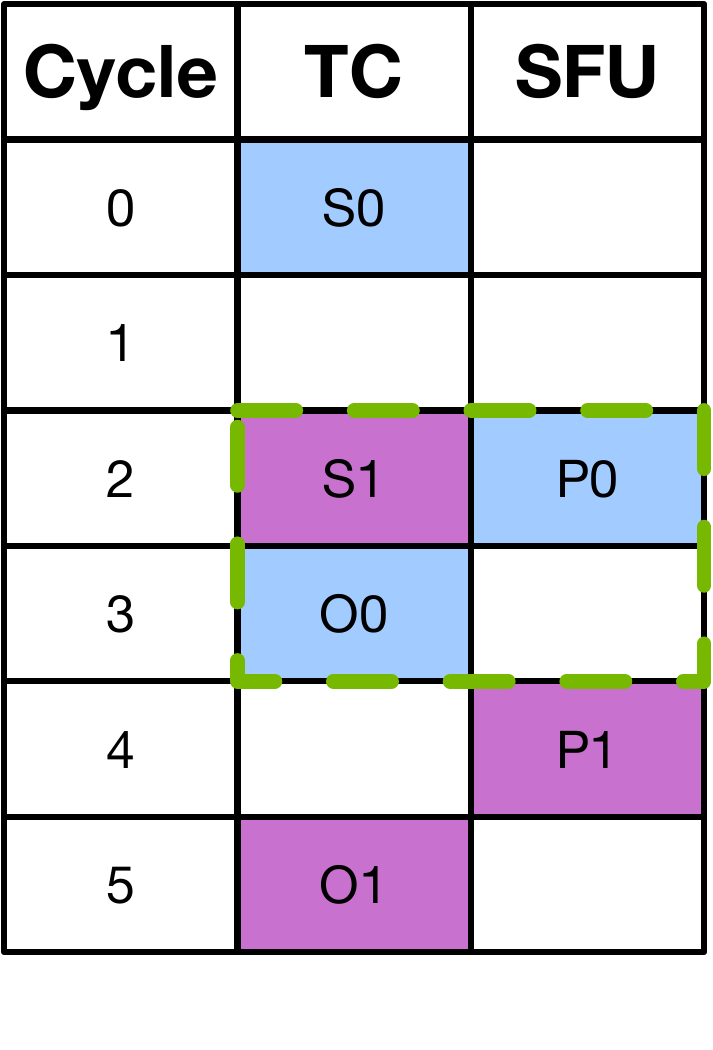}
\caption{$\loverii{}$ copies of \Cref{fig:attn-modulo-schedule}'s schedule placed $I$ cycles apart.}
\label{fig:attn-placed}
\end{subfigure}\hfill%
\begin{subfigure}[b]{0.235\textwidth}
\centering
\begin{lstlisting}
O = zeros()
// Prologue.
S = gemm(Q, K[0])
// Steady State.
for i:
  Sn = gemm(Q, K[i])
  P = exp(S)
  O += gemm(P, V[i-1])
  S = Sn
// Epilogue.
P = exp(S)
O += gemm(P, V[n-1])
\end{lstlisting}
\caption{Code generated from \Cref{fig:attn-placed} with a steady-state loop.}
\label{fig:attn-codegen}
\end{subfigure}\hfill%
\begin{subfigure}[b]{0.235\textwidth}
    \centering
    \includegraphics[width=0.65\textwidth]{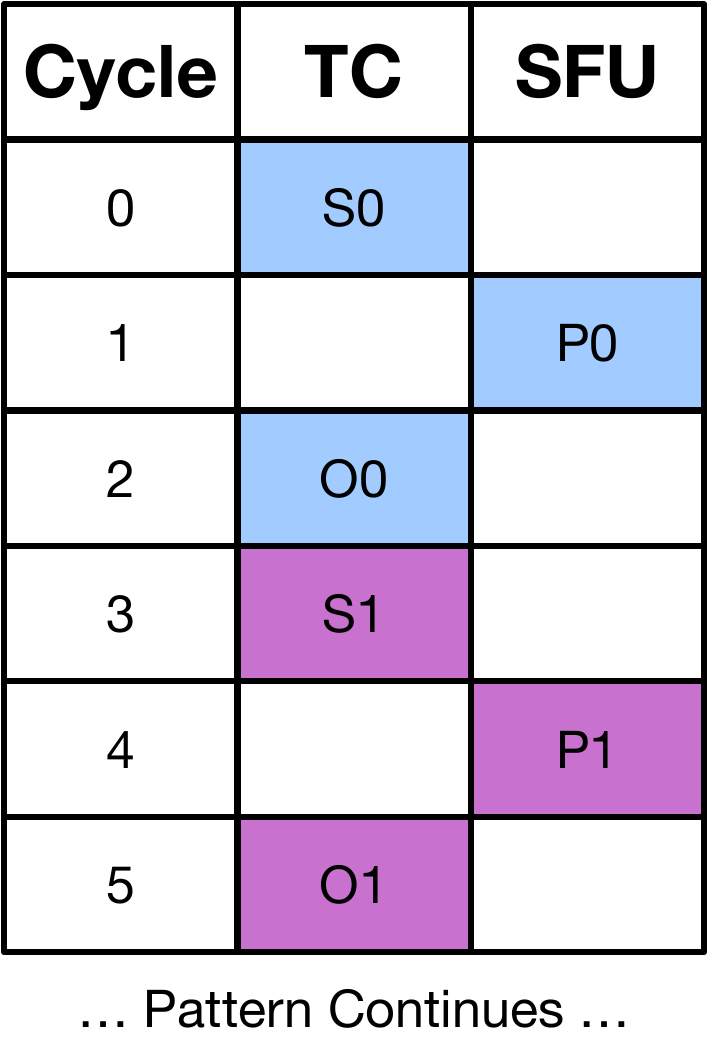}
    \caption{In-order execution, achieving $1/3$ its. per cycle.}
    \label{fig:attn-naive-exec}
\end{subfigure}\hfill%
\begin{subfigure}[b]{0.235\textwidth}
\centering
\includegraphics[width=0.65\textwidth]{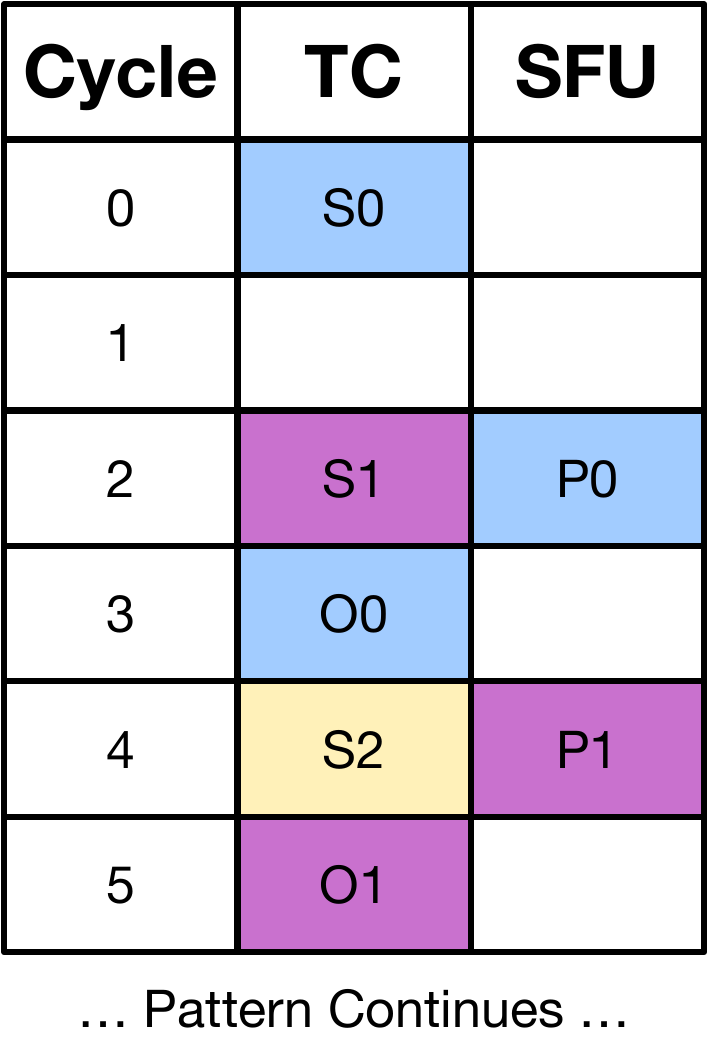}
\caption{Pipelined execution, achieving $1/2$ its. per cycle after prologue.}
\label{fig:attn-ppl-exec}
\end{subfigure}

\caption{Modulo scheduling a simplified Flash Attention expressed in a tile-based manner. The
machine costs are for Hopper, where \texttt{GEMM} and \texttt{EXP} on a tile have roughly the same cost. Modulo scheduling recovers the Flash Attention 3~\cite{flash-attention-3} pipeline.}
\label{fig:software-pipelining-example}
\end{figure*}

Software pipelining (SWP)~\cite{aiken-pipelining} reorders the operations
in a sequential program to maximize utilization of the
available functional units by exploiting instruction-level parallelism
(ILP) within and across loop iterations.
We use modulo scheduling~\cite{lam-modulo-scheduling, rau-modulo-scheduling}
to compute pipelined SWP schedules, and \Cref{sec:modulo-scheduling} introduces standard terminology using the running example shown
in \Cref{fig:software-pipelining-example}.  The figure shows how to
pipeline a simplified version of Flash Attention~\cite{flash-attention}.
The code (\Cref{fig:simpl-attn}), instruction cost information (\Cref{fig:attn-rrts}), and a dependency graph (\Cref{fig:attn-depgraph}) go through two 
intermediate steps (\Cref{fig:attn-modulo-schedule,fig:attn-placed}) before yielding the
pipelined code in \Cref{fig:attn-codegen}.
In this example, SWP is necessary for maximal TC
utilization because the program exposes the latency of the
exponential due to the data dependence on the result prior to
the second matrix multiplication (\Cref{fig:attn-naive-exec}).
%
However, as we discuss in \Cref{sec:code-generation}, it may not be possible on recent architectures for a single thread of execution to manifest this pipelined schedule. 
%
To address this issue, we describe in \Cref{sec:warp-specialization} how to realize an optimal pipelined schedule by 
distributing operations across multiple executing threads using WS, and the compilation challenges it induces. 

\subsection{Modulo Scheduling}
\label{sec:modulo-scheduling}

Modulo scheduling~\cite{lam-modulo-scheduling, rau-modulo-scheduling, gao-ilp-scheduling} is popular class of 
algorithms for SWP that transforms a given loop such that it achieves maximum throughput
(iterations finishing per unit time) while simultaneously abiding by the data
dependencies of the input program and capacities of the underlying machine.
%
In modulo scheduling, the input loop is described as a dependence graph
$G = (V, E)$ where $V$ is the set of instructions in the loop and $E$ defines
dependencies between the instructions.
We assume that $G$ describes a singly-nested loop without control flow.
%
In \name{}, each instruction represents a tile-level operation that
utilizes resources from the entire SM; dependence graphs of this form
can be extracted from standard tile-based programming models (see \Cref{sec:implementation}).

Associated with each $v \in V$ is a \emph{resource reservation table} (RRT)
that indicates the functional unit(s) used during the execution of $v$.
$\text{RRT}[v]$ is a 2-D integer array where each column corresponds to a kind of
functional unit, each row corresponds to a clock cycle in the execution of $v$,
and the entry gives the number of instances of $f$ that are occupied by $v$ at
that clock cycle of its execution.
The total number of instances of $f$ on the machine, called the capacity of $f$, is given by a machine description $D$.
RRTs for each operation in the running example and a machine description 
$D$ are shown in \Cref{fig:attn-rrts}; both instructions take a single cycle
to execute, but they occupy different functional units.
%
%
For simplicity we assume each functional unit can be scheduled and execute an instruction every cycle; on a real machine, functional units often take longer to execute some instructions than others and have limits on how often they accept new instructions.
For TC GPUs that operate on large tiles of data, the abstraction of an RRT is still valid at a coarser granularity of computation, where a ``cycle'' might represent a coarser granularity of time.
We describe how to cope with the problem of determining time granularity in \Cref{sec:cost-normalization}.

%
Every edge $e \in E$ is  a tuple $(u, v, d, \delta)$, where
$u$ is the source and $v$ is the sink of the data dependence.
The \emph{clock cycle delay} $d \geq 0$ indicates that $v$ must be issued at least
$d$ cycles after $u$ is issued.
Finally, $\delta \geq 0$ is the \emph{iteration delay}, indicating that
the instance of $v$ from loop iteration $i$ must be issued at least
$d$ cycles after the instance of $u$ from iteration $i - \delta$.
For example, an edge with $\delta = 1$ indicates that the dependence between
$u$ and $v$ is \emph{loop carried} to the very next iteration.
%
%
The dependence graph for the attention example is shown in
\Cref{fig:attn-depgraph}, where the edges from $S$ and $P$ have $d=1$ (from the
corresponding RRTs) and $\delta = 0$, while the loop-carried dependence on $O$
has $\delta = 1$.
To simplify the exposition, we use variable names (which are unique) instead of instruction names in the figures.
Together, the RRTs of each instruction (\Cref{fig:attn-rrts}) and the dependence graph $G$
(\Cref{fig:attn-depgraph}) are the inputs to modulo scheduling.

When run on a given loop, modulo scheduling finds an \emph{initiation interval}
$I$ and \emph{modulo schedule} $M$, from which the pipelined loop is constructed.
%
%
$I$ corresponds to the rate at which new loop iterations are issued (or
finished), and is thus inversely related to the throughput; $I=1$ is the minimum
possible and means that a new loop iteration is launched every cycle.
$M$ maps each instruction to the clock cycle at which it must be issued,
counting from the beginning of its iteration (hence the name ``modulo'').
If $M(v) = k$, then $v$ will be issued at clock cycles $k, I + k, 2I + k, \dots$ in the pipelined loop.
A choice of $M$ and $I$ is valid when it ensures that the pipelined loop
satisfies all dependence edges in $E$ and keeps all functional units within
capacity.
%
%

A valid modulo schedule for the running
example is shown in \Cref{fig:attn-modulo-schedule}.
It is accompanied by the \emph{modular RRT}, a data structure similar to the 
RRT that indicates
the functional unit usage in the steady state of the pipelined loop.
In the modulo schedule, both $P$ and $O$ were delayed
one cycle later than an optimal schedule for just a single iteration.
$M(P) = 1$ and $M(O) = 2$ would be valid per the data dependencies but invalid per the modulo RRT because $O$ would conflict with $S$ for the TC since $0 \equiv 2 \; (\operatorname{mod} 2)$.
A variety of algorithms have been developed to construct
$M$ and $I$ for a given dependence graph $G$, including
greedy~\cite{lam-modulo-scheduling, rau-modulo-scheduling} algorithms
and optimal~\cite{gao-ilp-scheduling, stoutchinin-ilp} algorithms
that leverage Integer Linear Programming (\Ilp{}).
%
Optimal algorithms yield modulo schedules with the smallest possible $I$,
and thus the highest possible throughput.


Once a modulo schedule $M$ and initiation interval $I$ have been found, they can be used to synthesize a program that executes the schedule. 
%
%
A standard way~\cite{dragon-book} to realize a modulo schedule is with a sequential loop that consists of: 1)
a \emph{prologue} that primes the loop, 2) a \emph{steady state}
containing the repeatedly executing loop body, and 3)
a \emph{epilogue} that drains the loop.
To construct these three components, we let $L$ be the \emph{length}
of $M$, i.e., the number of cycles in $M$.
The software-pipelined loop is constructed by overlapping $\left\lceil L/I \right\rceil$
copies of $M$, each copy offset by $I$ cycles from the previous copy.
The pipelined loop can then be read off from this staggered schedule: 
the first $ \left(\left\lceil L/I \right\rceil - 1 \right) \cdot I$ cycles correspond to the prologue,
the next $I$ cycles correspond to the looping steady state, and the 
remaining cycles are the epilogue that drains the pipeline.
Placing the modulo schedule for the running attention example is shown in \Cref{fig:attn-placed},
where blue and purple correspond to different copies of the schedule, and the dashed box gives the steady state.
Once the schedule has been constructed, it can be used to generate code.
Generated pseudocode for the running example is shown in \Cref{fig:attn-codegen}.
An in-order execution as written in \Cref{fig:simpl-attn}
completes 1 loop iteration every 3 cycles (\Cref{fig:attn-naive-exec}), while
the pipelined execution completes 1 iteration every 2 cycles after the prologue (\Cref{fig:attn-ppl-exec}).

Applying modulo scheduling to the running Flash Attention example
yields the exact pipeline developed by experts in Flash Attention 3~\cite{flash-attention-3}
from only a high-level description of the algorithm and machine.


%
%
%

\subsection{Code Generation Challenges}
\label{sec:code-generation}

For GPUs prior to Hopper, the modulo scheduled tile-level program in \Cref{fig:attn-codegen} could be lowered to single-threaded code
executing in a SIMT fashion, where each data-parallel thread would execute the same program on disjoint subsets of the tile of data.
%
%
%
Unfortunately, most modulo schedules computed for the functional units of a Hopper or a Blackwell GPU cannot be realized as a single-threaded program in the standard style.
There are four confounding factors that regularly thwart single-threaded code generation.
First, and most commonly, the scale of TCs in recent GPUs  requires multiple warps to cooperatively issue large GEMM computations.
Most GPU compilers are built on sequential intermediate representations (e.g., LLVM and PTX) and are therefore ill-equipped to generate code for cooperative warps. 

A second issue stems from the propensity of modulo schedules to increase the working set size of computations to keep live variables from multiple iterations of a program on-chip concurrently.
The combination of this natural memory pressure of modulo schedules in conjunction with the larger quantities of data needed to feed bigger TCs makes it challenging to keep the working set of a computation on-chip.
Most GPUs have a strict upper bound of 255 registers per thread and spilling data out of the register file often incurs a significant performance penalty as the spilled data is unlikely to remain in cache.
Many single-threaded realizations of modulo schedules for recent GPUs struggle to achieve peak performance due to the cost of spilling.

A third complication is that some operations, such as transfers through
the multi-tiered GPU memory hierarchy, have high
variability in their execution times.
While traditional modulo scheduling suggests using an upper bound 
on such costs~\cite{dragon-book}, there is an implicit assumption that this upper 
bound is roughly the same order of magnitude as other operations in the schedule.
However, operations like TMA transfers that move tiles of
data between global and shared memory may have 
more than an order of magnitude difference
between the fastest and slowest possible execution times for the same transfer.
%
While these operations are performed asynchronously, the high dynamic range of 
potential latencies can foil attempts to place synchronization instructions during code generation.
In particular, overestimation results in under-utilization, while
underestimation results in pipeline stalls during execution.

\begin{figure}
    \centering
    \includegraphics[width=0.75\linewidth]{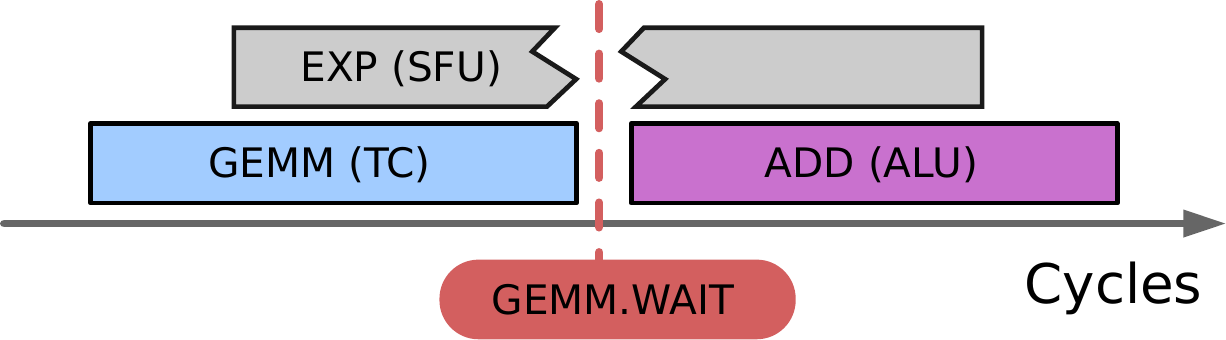}
    \caption{Visualization of three operations using
    different functional units scheduled on the same warp. The blocking sync
    after \texttt{GEMM} interrupts the concurrent issue of \texttt{EXP}.}
    
    \label{fig:blocking-sync-example}
\end{figure}

Finally, the fixed-function units on Hopper and Blackwell have 
asynchronous interfaces that require explicit blocking synchronization to consume results.
Since GPU threads have \emph{in-order} instruction issue, blocking synchronization
interrupts the instruction issue of any concurrently scheduled operations
on the same warp.
An example is in \Cref{fig:blocking-sync-example}, where a modulo schedule
requires an addition to consume the result of a \texttt{GEMM}, while executing an independent
{\tt EXP} concurrently.
Note that all three occupy independent functional units.
Executing all operations from the same warp suffers degraded performance
as the \texttt{GEMM.WAIT} interrupts instruction issue for \texttt{EXP}.
%
%
%
%
%
%

Each of these four factors influence code generation for modulo schedules to differing degrees depending on the properties of the particular program.
Combating them in practice necessitates a new approach to code generation.

\subsection{Warp Specialization}
\label{sec:warp-specialization}

Warp specialization~\cite{cuda-dma, singe} (WS) is a programming paradigm that offers a solution to the code generation challenges encountered when trying to realize a modulo schedule on GPUs.
A warp-specialized program assigns operations to different warps while ensuring that the warps cooperatively execute the entire computation.
WS is possible due to the hierarchical grouping of threads within an SM.
Since warps are SIMT, a warp's performance is maximized when all threads in the warp issue the same instruction, and minimized when the threads \emph{diverge}.
However, unlike threads, warps are independent from each other and pay no penalty for control divergence.

While WS is commonly viewed as a separate optimization, our insight is that
WS is useful precisely because it directly addresses all of the code generation
problems for SWP presented in \Cref{sec:code-generation}.
%
First, WS naturally reasons about warps cooperating and can ensure that subsets of warps cooperatively issue TC operations.
Second, by splitting the computation across many warps, the computation can access the register resources of many threads, yielding greater flexibility for fitting the large working sets demanded by SWP on-chip.
Third, variable-latency operations can be separated onto dedicated warps, instead of
needing to be statically scheduled alongside (and interfering with) fixed-latency operations.
Finally, splitting operations across warps allows some warps to issue instructions
while others are blocked on synchronization.

Since WS addresses the challenges of generating code for SWP, it has gained
significant adoption by both compilers~\cite{triton, tawa, cypress, pallas, tile-lang, tlx} and programmers~\cite{flash-attention-3, flash-attention-4, cutlass}.
%
However, there are two complications.
First, WS is not free and comes with trade-offs involving communication of data between warps and synchronizing access to both shared data and hardware resources. 
Currently, all automated approaches that we are aware of depend upon ad-hoc heuristics to decide these trade-offs~\cite{triton, tawa, cypress}.
These heuristics often involve canonical warp ``roles'' (e.g., loader and compute warps), or treat warps as parallel ``agents'' to whom work can be dispatched.
Without a technical framework for understanding the optimality of these heuristics, it is difficult to know if programs generated using them are achieving peak performance.

The other significant complication to this approach is that, even with WS, it might not be possible to synthesize a program that actually achieves a modulo schedule.
For example, despite spreading a computation across as many warps as possible in an SM, it still might be impossible to fit the working set demanded by a modulo schedule in the register file.
Consequently, treating the computation of the modulo schedule and code generation using WS as two independent steps in compiling a program can lead to sub-optimal code in practice.
To ensure that a modulo schedule can actually be realized, the constraints for code generation using WS need to be directly incorporated into the optimization process in conjunction with those for traditional modulo scheduling.

\section{Joint Optimization Problem}\label{sec:joint-problem}

We now demonstrate how to solve the optimization problem of finding a modulo schedule and a WS strategy simultaneously.  
The joint optimization problem is to take an input program $G = (V, E)$ and produce a modulo schedule with the minimum initiation interval along with a WS strategy capable of realizing the schedule.
The result is a modulo schedule $M^*$, initiation interval $I^*$ and warp assignment $A^*$, where $A^*$ is an assignment of every $v \in V$ to a warp (or warps).

We approach this problem in two main steps.
First, modulo scheduling is used to derive an 
initial modulo schedule $M$ and initiation interval $I$
that achieve the maximum throughput while respecting the data dependence and
functional unit constraints.
Then, $M$ and $I$ are used to seed a system of constraints that are supplemented 
with constraints for WS.
The system of constraints can then be solved by an SMT solver to discover 
an $M^*$ and $A^*$ with the same $I$
that respect the requirements of both modulo scheduling and WS.

\begin{figure}[t]
        \centering
        \includegraphics[width=\linewidth]{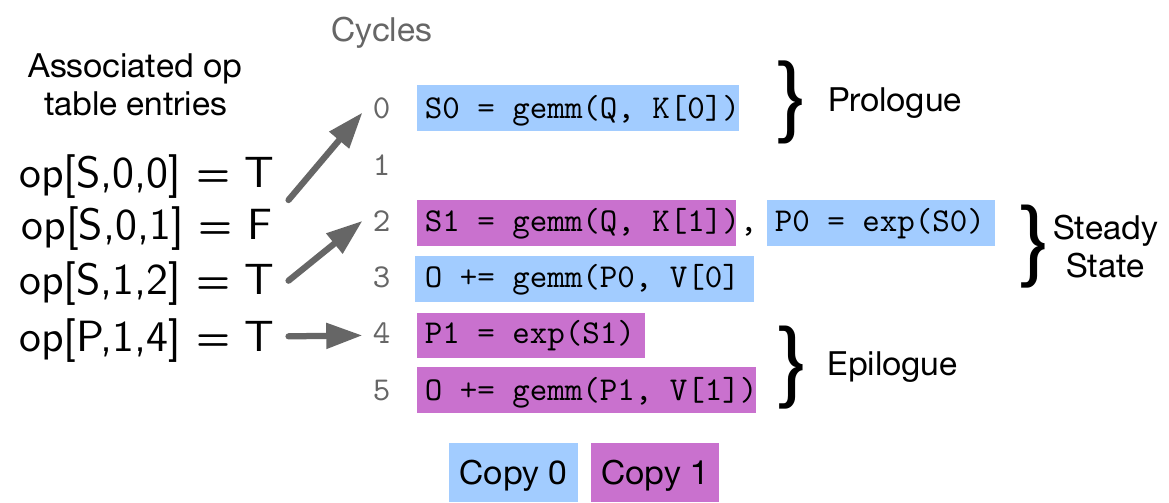}
        \caption{Straight-line code analyzed by \name{}'s joint formulation. Sample $\operatorname{op}$ table entries are to the left.}
        \label{fig:smt-straightline-code}
\end{figure}

We use $M$ and $I$ to define an initial
straight-line program $Q$ over which the 
constraint system is formulated.
Q is obtained using the code-generation procedure of
modulo scheduling from \Cref{sec:code-generation}.
Recall that $\loverii{}$ copies of $M$ are overlapped, each 
offset $I$ cycles from the previous,
resulting in a prologue, steady state and epilogue.
Our insight is that we can assume that the steady state
executes exactly once, allowing us to treat the three parts
as a straight-line program with length $T$.
This assumption is sound, because it is permissible for the steady state to execute exactly once, and complete, because steady state executions are identical, and greatly simplifies the constraints by obviating the need for any loop analysis.
An example program $Q$ derived from the modulo scheduled example
in \Cref{fig:software-pipelining-example} is shown in the right
half of \Cref{fig:smt-straightline-code}.

\subsection{Modulo Scheduling with Constraints}\label{sec:constraints-modsched}

\begin{figure}[t]
    \small
    \begin{align*}
      &\forall v, i \quad \sum_{t} \operatorname{op}[v, i, t] = 1 \tag{\textsc{Uniqueness}}\\
      &\forall v, i \in\left[1, \lceil L/I \rceil \right), t \quad \operatorname{op}[v, 0, t] \Rightarrow \operatorname{op}[v, i, t + i \cdot I] \tag{\textsc{Consistency}} \\
      &\forall v, i, t \quad t + \text{cycles}(v) > T \Rightarrow \neg \operatorname{op}[v, i, t] \tag{\textsc{Completion}} \\
      &\forall i, t, (u, v, d, \delta) \in E, ~t' \in [0, t + d) \quad \operatorname{op}[u, i, t] \Rightarrow \neg \operatorname{op}[v, i + \delta, t'] \tag{\textsc{Dependence}}\\
      &\begin{aligned}
      &\forall t, f \quad \sum_{\mathclap{v, i, c \in \left[0, \text{cycles(v)}\right)}} \operatorname{op}[v, i, t - c] \cdot \operatorname{RRT}[v][f, c] \leq \text{cap}(f)
      \end{aligned} \tag{\textsc{Capacity}}
    \end{align*}
    \caption{Constraints enforcing a valid modulo schedule.}
    \label{fig:modulo-scheduling-constraints}
\end{figure}

The goal of the joint problem is to find a modulo schedule $M^*$ that may be distinct from $M$ but achieves the same initiation interval. 
To do so, \name{} must be able to modify $M$, necessitating a \emph{reimplementation} of modulo scheduling in the form of SMT constraints.
Instead of modifying $M$ directly, this reimplementation allows the modification of $Q$ to $Q^*$ such that $Q^*$ has the same initiation interval and length as $Q$, but may be the result of a different modulo schedule.
The constraints to do so are given in \Cref{fig:modulo-scheduling-constraints}.
We define a 3-D boolean array such that $\text{op}[v, i, t]$ indicates that operation $v \in V$ of iteration $i \in [0, \loverii{})$ was scheduled at clock cycle $t \in [0, T)$ in $Q^*$.
The uniqueness constraint enforces that every operation is scheduled exactly once.
The consistency constraint ensures that $Q^*$ is obtainable from some modulo schedule.
The completion constraint requires all operations to finish before the last clock cycle $T$ of $Q^*$.
Finally, the dependence and capacity constraints function identically to the corresponding constraints in modulo scheduling.
We elide guards for brevity, but assume that constraints referencing $i \not\in[0, \loverii{})$
are not emitted.
Additionally, references to $t \not\in[0, T)$ of op are mapped to false.
A satisfying assignment of op induces a valid $Q^*$ and $M^*$, where $M^*(v) = t$ iff op$[v, 0, t]$.

\subsection{Memory Aware Constraints}\label{sec:constraints-mem}

\begin{figure}[t]
    \small
\begin{align*}
&\forall t, m \quad \sum_{v, i} \text{live}[v, i, t] \cdot \text{footprint}(v, m) \leq \text{capacity}(m) \tag{\textsc{Memory Capacity}}\\
&\forall v \quad \left(\exists (v, u, d, \delta > 0) \in E\right) \Leftrightarrow \text{live}[v, \lceil L/I \rceil -1, T] \tag{\textsc{Init}}\\
&\forall v, i, t \quad (\text{live}[v, i, t] \wedge \text{op}[v, i, t]) \Rightarrow \neg \text{live}[v, i, t-1] \tag{\textsc{LiveProp-1}}\\
&\forall v, i, t \quad (\text{live}[v, i, t] \wedge \neg \text{op}[v, i, t]) \Rightarrow \text{live}[v, i , t-1] \tag{\textsc{LiveProp-2}}\\
&\forall v, i, t \quad (\neg \text{live}[v, i, t] \wedge \bigvee_{\mathclap{(v, u, \_, \delta) \in E}} \text{op}[u, i + \delta, t]) \Rightarrow \text{live}[v, i, t-1] \tag{\textsc{DeadProp-1}}\\
&\forall v, i, t \quad (\neg \text{live}[v, i, t] \wedge \bigwedge_{\mathclap{(v, u, \_, \delta) \in E}} \neg \text{op}[u, i + \delta, t]) \Rightarrow \neg \text{live}[v, i, t-1]\tag{\textsc{DeadProp-2}}
\end{align*}
\caption{Memory Allocation Constraints.}
\label{fig:mem-alloc-constraints}
\end{figure}

To ensure that the working set of a modulo schedule can remain on-chip, we introduce additional constraints that enforce memory capacity,
narrowing the set of valid instances of $Q^*$.
These constraints assume that $Q^*$ is in static single assignment form (SSA) \cite{cytron-ssa, appel-compilers},
where variables are defined once and have lifetimes that end at the last use.
We define a second boolean array $\text{live}[v, i, t]$, where $\text{live}[v, i, t] = 1$
when the result of the $i$'th instance of $v$ is live at time $t$.
Using live, defining the memory capacity constraint is straightforward and shown in \Cref{fig:mem-alloc-constraints}, where the memory footprints and capacities are defined
by the input graph and machine model.
The non-trivial component of memory capacity is setting up the constraints
defining $\text{live}[v, i, t]$; this component cannot be discharged to an external analysis, because
changes to $Q^*$ affect when different values are live.
These constraints are shown in \Cref{fig:mem-alloc-constraints}, and are derived from classic
backwards dataflow algorithms for liveness \cite{dragon-book}.
There are three groups of liveness constraints: initial conditions, propagation of liveness, and
propagation of deadness.
%
%
Initially, only the loop-carried results of the last copy of each instruction
are live at the last clock cycle of $Q^*$, i.e. at time $T$.
%
%
Liveness is then propagated backwards through the straight-line program: if an operation's
result is live at time $t$, it is live at time $t-1$ unless it was scheduled at time $t$.
Deadness is propagated in a similar way, where if the result of an operation is dead
at time $t$, it stays dead at time $t-1$ unless a user of the operation is scheduled
at time $t$.
%

\subsection{Warp Assignment Constraints}\label{sec:constraints-ws}

\begin{figure}[t]
    \small
\begin{align*}
&\forall v \quad \sum_{w} \text{opw}[v, w] = 1 \tag{\textsc{Warp Uniqueness}}\\
&\forall v \quad \text{variable\_latency}(v) \Leftrightarrow \text{opw}[v, W_{vl}]\tag{\textsc{Variable Latency}}\\
&\forall t, w \quad \sum_{v, i} \text{live}[v, i, t] \cdot \text{opw}[v, w] \cdot \text{regs}(v) \leq \text{reg\_limit}() \tag{\textsc{Register Limit}}\\
&\begin{aligned}
&\forall (u, v, d, \delta) \in E, t, i, w, w' \neq w, s \in [0, \text{spillcost}(u)) \\
&\quad \text{op}[u, i, t] \wedge \text{opw}[u, w] \wedge \text{opw}[v, w'] \Rightarrow \\
& \quad \neg \text{op}[v, i + \delta, t + d + s] \\
\end{aligned} \tag{\textsc{Cross-Warp Spills}} \\
&\begin{aligned}
&\forall (u, v, \_, \_) \in E, t, w, i, o\neq v \\
& \quad \text{op}[v, i, t] \wedge \text{opw}[v, w] \wedge \text{blocking}(u, v) \Rightarrow \\ 
& \quad \forall i', t' \in [t - (\text{cycles}(o) - 1), t], ~ \neg (\text{op}[o, i', t'] \wedge \text{opw}[o, w])
\end{aligned} \tag{\textsc{Concurrency}}
\end{align*}
\caption{Warp Assignment Constraints.}
\label{fig:warp-assignment-constraints}
\end{figure}

Finally, we introduce constraints that influence WS decisions, 
shown in \Cref{fig:warp-assignment-constraints}.
We define another boolean array $\text{opw}[v, w]$, indicating whether
operation $v$ has been assigned to warp $w$;
opw directly defines the final warp assignments $A^*$.
Similar to op, we define a uniqueness constraint,
stating that every operation must be assigned to exactly one warp.
This constraint is naturally extended to operations that span multiple warps, such as
the warp-group level operations on Hopper and Blackwell.
%

The first constraint is the assignment
of operations with variable and statically-unknown latencies.
These operations are assigned to a designated warp $W_{vl}$,
separating them from the scheduling and placement of 
instructions with statically-known and fixed latencies.
The second constraint enforces the register limit on each warp;
the constraint is similar to the capacity constraint in \Cref{fig:mem-alloc-constraints},
but scopes the summation to each warp.
This constraint forces a partitioning of $V$ to stay within register limits.
However, data in the registers of warp $w$ cannot be accessed on
warp $w'$ unless the data is transferred (or \emph{spilled}) through the shared memory.
When data is spilled across warps, a valid schedule must include the cost of
the transfer when scheduling the consuming operation; this additional
delay is captured by the \emph{cross-warp spill} constraint.
We assume the cost of cross-warp communication is provided as an annotation
on nodes in $G$.
This constraint forces a trade-off between memory capacity, latency and
the costs of cross-warp spills.
The final constraint is the \emph{concurrency} constraint, 
which captures the effects of blocking synchronization interrupting the issue
of concurrently scheduled operations.
Certain edges $(u,v)$ are designated as \emph{blocking} edges, indicating
that blocking synchronization is required by $v$ before the results of $u$ are accessible.
The concurrency constraint states that if an operation $v$, assigned to warp $w$
and scheduled at time $t$, requires blocking synchronization from an incoming
dependence, then no other operation can be scheduled to run on $w$ when $v$ starts;
this forces concurrent operations to either be placed onto different warps or scheduled
at different times.
While elided for brevity, the implementation includes a similar constraint for cross-warp spills,
as transferring register data through shared memory requires blocking synchronization.

The entire system of constraints captures all the challenges presented in \Cref{sec:scheduling} that can make a desired modulo schedule impossible to implement in practice for recent GPUs.
By describing the space of solutions holistically with a unified system of general constraints, we force the constraint solver to reckon with all aspects of the scheduling problem simultaneously.
Furthermore, as new machines are released, constraints can easily be added, removed, and modified to reflect the challenges presented in future architectures.
Therefore our approach provides a flexible framework for finding optimal solutions to the joint problems of modulo scheduling and WS.

\section{Implementation}\label{sec:implementation}

To reify our approach,
we introduce \name{}, a system that computes joint 
SWP and WS strategies for Triton~\cite{triton} programs.
\name{} extracts dependence graphs from a mid-level Triton intermediate representation (IR) called TTGIR, which is a tile-based, SSA IR with arithmetic on tiles and 
explicit data movement.
%
These graphs serve as the input program to \name{}'s optimization process as described in \Cref{sec:joint-problem}.
%
%
%
Users of \name{} must also inform it of the target GPU architecture.
The target architecture is used to estimate costs of instructions and data movement
(discoverable via documentation~\cite{cuda-tp-docs} or direct measurement),
and to enable architecture-specific modeling, such as declaring available memories or
denoting operations that require blocking synchronization.


As described in \Cref{sec:joint-problem}, \name{} computes an 
initial modulo schedule by formulating it as an \Ilp{} problem \cite{stoutchinin-ilp} and 
dispatching it to the CBC solver~\cite{cbc-solver}.  
\name{} then uses the resulting modulo schedule and initiation interval to seed the system
of SMT constraints.
\name{} discharges SMT queries to the quantifier-free linear integer
arithmetic (QFLIA) theory of the Yices2 SMT solver~\cite{yices}.
%
Once a modulo schedule and warp assignment have been discovered,
\name{} uses standard code generation techniques to emit software-pipelined IR, 
annotating each instruction with the warp (or warps) that should execute it.
The resulting IR can either be consumed by downstream compilers
that implement a specified WS strategy (e.g., 
Tawa~\cite{tawa} or Cypress~\cite{cypress}), or used by an expert
as reference for a manual implementation.

\subsection{Handling Unsatisfiability}

\name{} must handle the case when the constraint system
is unsatisfiable.
Unsatisfiability indicates that some resource limitation
or concurrent interaction is preventing a modulo schedule
with the initial initiation interval $I$ from 
being achieved.
Similar to standard approaches in modulo scheduling~\cite{lam-modulo-scheduling}, if the constraints are unsatisfiable,
\name{} uses modulo scheduling to find a new candidate schedule $M'$ with
initiation interval $I+1$, which is then used to seed the system of constraints for another solution attempt.
By searching monotonically from the smallest possible initiation interval,
we ensure discovery of the highest throughput schedule that solves the constraints.
In addition to searching over candidate values of $I$, \name{} also searches over
increasing values of $L$ (the total schedule length) that do not affect $\loverii{}$ at a given $I$.
This search is depicted in \Cref{alg:workflow}.

\begin{algorithm}[t]
\caption{\name{}'s Search Procedure}
\label{alg:workflow}
\small
\begin{algorithmic}[1]
    \Procedure{\name{}}{$G$}
    \State $I \gets 0$
    \While{\textbf{true}}
        \State $I \gets I+1$
        \State $M \gets \Call{Optimal-Modulo-Schedule}{G, I}$
        \If{$M = \textbf{failure}$}
            \State \textbf{continue}
        \EndIf
        \State $L \gets \Call{Len}{M}$
        \While{$\lceil L/I \rceil = \lceil \Call{Len}{M} / I \rceil$}
            \State $(M^*, A^*) \gets \Call{SWP-and-WS}{G, M, I, L}$
            \If{$(M^*, A^*) = \textbf{failure}$}
                \State $L \gets L + 1$
                \State \textbf{continue}
            \EndIf
            \State \Return $(M^*, I, A^*)$
        \EndWhile
    \EndWhile
    \EndProcedure
\end{algorithmic}
\end{algorithm}

\subsection{Cost Normalization}
\label{sec:cost-normalization}

A critical component of \name{}'s implementation is \emph{cost normalization},
which renders the optimization problems solved by \name{} tractable.
%
Consider the cycle counts computed from publicly available
documentation~\cite{cuda-tp-docs} for a 128x128x128 GEMM on Hopper---roughly 1000 cycles.
A reasonable dependence graph $G = (V, E)$ may contain many instructions with similar cycle counts.
Optimal algorithms for modulo scheduling and \name{}'s joint formulation
are not just exponential in $|G|$, but exponential in $\sum_{(u, v, d, \delta) \in E} d$.
Therefore, directly using estimated cycle counts results in intractable \Ilp{} and SMT problems.


Our solution relies on the observation that multiplying every cycle count in a
SWP problem by a positive integer results in a new problem
isomorphic to the original.
It is not the cycle counts that matter, but the \emph{ratios} between
cycle counts.
%
%
Therefore, we want to obtain new, smaller cycle counts whose ratios are as close as possible to ratios of the original cycle counts.
Cost normalization encodes this intuition as an \Ilp{} problem (separate from the one for modulo scheduling).



Let the list of integers $C$ be the original cycle counts.
We want to derive a new list of integers $C'$ such that $C[i] / C[j] \approx C'[i] / C'[j]$ for all $i, j$.
%
This is formalized as a constraint by introducing a variable $F$ that bounds the change in ratios.
We also bound the sum of $C'$ from above and below, the latter to avoid the degenerate solution of all zeros:
{
\small
\[
   \forall i,j \quad  -F \leq C[i]\cdot C'[j] - C[j] \cdot C'[i] \leq F
\]
\[
  1 \leq \sum_i C'[i] \leq U
\]
}
$U$ is a user-defined integer parameter that controls the exact trade-off between resolution of costs and running time of the subsequent algorithm that actually calculates a modulo schedule.
Smaller values of $U$ result in lower resolution (i.e., larger change in ratios) and lower running times.
%
%
Finally, the \Ilp{} objective is to minimize $F$.
\name{} uses the SCIP solver~\cite{scip-solver} to find solutions because it was considerably faster than the CBC solver~\cite{cbc-solver} at such problems.
In our experiments, we pick $U = 300$ and find that SCIP is able to find the global minima in under 500 ms in all cases.
Early in the development of \name{}, we relied on an ad-hoc approach that
divided each cycle count by a fixed integer, followed by rounding to obtain
integer values.
We found it difficult to keep this approach consistent,
even for the same program on different GPUs.
In contrast, framing cost normalization as an \Ilp{} problem offers a principled solution and is
necessary for discovering optimal schedules.

\subsection{Variable Latency Optimizations}

As discussed in \Cref{sec:code-generation}, approximating the execution time
of variable-latency operations with a high dynamic range can cause imprecision
in static scheduling.
While \name{} offloads variable-latency operations onto separate
warps so they may be dynamically scheduled, further optimizations are possible
for variable-latency operations in the target loop dependence graph that
have no incoming data dependencies.
We refer to these variable-latency operations as \emph{streaming} operations.
On separate warps, streaming operations can run ahead of the main software pipeline 
and complete several iterations before their results are needed.
To reflect this, we assign streaming operations zero latency in our cost models,
allowing dependent operations with statically-known latencies to be precisely scheduled.
We then expose the pipeline depths for these streaming operations as
parameters that may be tuned by an external auto-tuning system; exposing
such parameters for dynamic tuning is commonplace among existing TC GPU programming
systems~\cite{cypress, cutlass, cutile, triton, pallas, thunderkittens}.
%
Many critical compute-bound kernels contain streaming variable-latency
operations (like TMA loads of input tiles), warranting optimization of this
important case.


\subsection{Limitations and Future Work}
Currently, our implementation of \name{} only supports singly-nested loops without additional control flow.
This limitation could be lifted through the use of hierarchical reduction techniques from the software pipelining literature \cite{lam-modulo-scheduling}, which we leave as future work.
While \name{} derives an optimal pipeline with respect to the constraints
presented so far, the tile size is not automatically determined by \name{}, and
must therefore be picked by a human or a higher-level auto-tuning system.
As we will see in \Cref{sec:evaluation}, the solution times of \name{}'s joint problem range from tens of seconds to a few minutes.
Fast solution times are a non-goal of our approach; we trade-off solution times for an optimality guarantee.
We envision a system like \name{} serving as a developer aid or as an offline compilation tool used prior to deployment, rather than being run continuously during interactive development.

\section{Evaluation}\label{sec:evaluation}

%
We do a thorough evaluation on two important kernels from the machine
learning literature, and show that \name{} is able to derive SWP and WS
strategies for \emph{multiple generations} of NVIDIA GPUs automatically and from
first principles.
\name{} is the only system we are aware of that is capable of doing
so without heuristics tailored for each GPU generation.
We then implement these strategies and show the achieved performance
is competitive with hand-tuned implementations.

\subsection{Methodology}

\paragraph{Kernel selection.}
We focus our evaluation efforts on compute-bound kernels that
require utilizing both the TC and general-purpose functional
units on the SM.
Hence, we focus on the critically important forward and backward
passes of Fused Multi-Head Attention, which have received
significant human effort in developing SWP and
WS strategies due to their importance in artificial intelligence systems~\cite{flash-attention-3, flash-attention-4}.
The backward pass requires a significantly different loop structure than the forward pass and thus demonstrates \name{}'s flexibility in discovering different strategies.


\paragraph{Evaluation Platforms.}
We evaluate \name{} on NVIDIA Hopper and Blackwell GPUs.
We use a NVIDIA H100 SXM5 80 GB and a NVIDIA B200 180 GB.
%
All experiments use CUDA 13.0.
Solution times for \name{}'s joint problem
were recorded on a single core of a Intel Xeon Platinum 8570.

\paragraph{Experimental Setup.}
\label{sec:exp-setup}

\name{} schedules dependence graphs extracted from Triton programs.
Initially, we intended to use \name{} within Triton and let 
Triton generate code from \name{}'s SWP and 
WS strategies.
However, there are many additional (and orthogonal to \name{}) 
critical decisions and optimizations that must be performed correctly to 
achieve high performance on modern GPUs.
We found that Triton made many incorrect decisions
during code generation, such as in memory allocation,
data layout conversions or synchronization placement.
%
%
As a result, Triton was either unable to successfully compile
\name{}'s pipelines, or resulted in poorly performing code.
To demonstrate the performance of schedules found by \name{},
we instead ``hand-compile'' \name{}'s pipelines into CUDA C++.
This process involved translating the pipelined and
warp-annotated IR emitted by \name{} into CUDA that implemented
the specified strategy while allowing us to correctly
make the remaining (and orthogonal) lowering steps.
Automating all of these remaining steps in an optimal manner requires
significant engineering and further research; as such, it is out of scope for
this work.

\subsection{Attention Forward Pass}

The forward pass of FMHA has been the focus of significant
optimization over multiple generations of GPU hardware
through the Flash Attention (FA) algorithms~\cite{flash-attention, flash-attention-2, flash-attention-3, flash-attention-4}.
We focus on FMHA on Hopper and Blackwell,
targeted by Flash Attention 3 (FA3)~\cite{flash-attention-3} and
Flash Attention 4 (FA4)~\cite{flash-attention-4} respectively.
For both architectures, novel SWP and
WS strategies were required for peak performance and
were discovered by humans months after each architecture was released.
We demonstrate that \name{} automatically rediscovers the same strategies
across different architecture generations.
%

We started with the Triton tutorial~\cite{triton-tutorial-attn} implementation
of FA, a high-level implementation of FA without additional optimizations.
%
A Triton program describes the decomposition of work for each SM.
For \name{}'s joint approach to SWP and WS,
we sub-tile the computation for the TC instruction
sizes of the target hardware.
This sub-tiling allows \name{} to control scheduling and warp assignments
at the appropriate level of granularity; Triton's IR only implicitly
reasons about the decomposition of tiles onto multiple warps.
We discuss these results in depth, focusing on each architecture in turn.


\subsubsection{Hopper}

\begin{figure}
\centering
\includegraphics[width=0.95\linewidth]{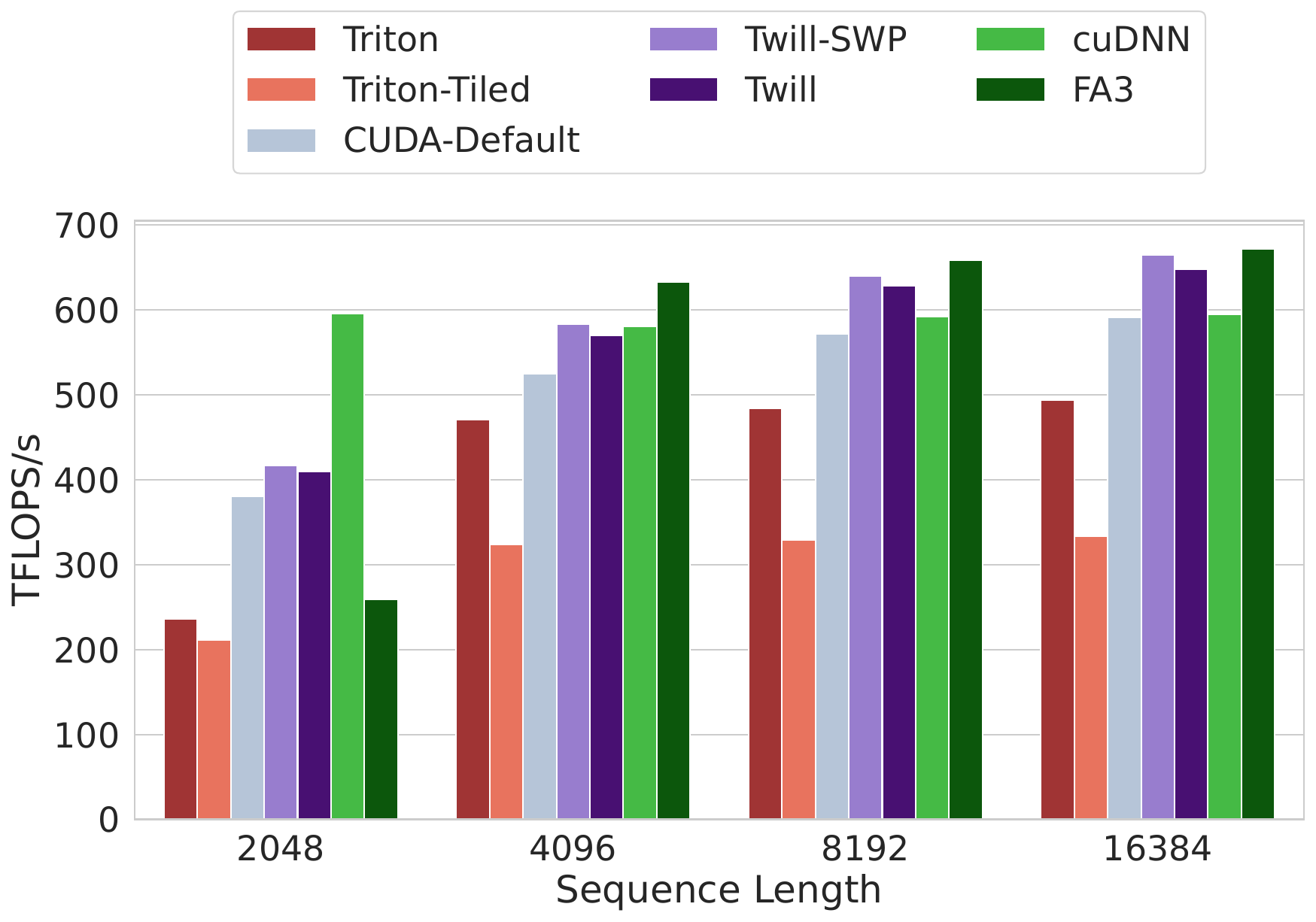}
\caption{Hopper FP16 Non-Causal Forward Attention. \texttt{(BATCH=4, NUM\_HEADS=32, HEAD\_DIM=128)}}
\label{fig:hopper-forwards-data}
\end{figure}

FA3~\cite{flash-attention-3} introduced two main techniques
for optimizing FMHA on Hopper.
The first technique was a software pipelining schedule for the main
loop that extracts one GEMM into the loop's prologue so that latency
of the exponential is not exposed (seen in \Cref{fig:software-pipelining-example}).
The second technique was called \emph{ping-pong scheduling}, which schedules the
GEMM from one sub-tile while the exponential from another sub-tile executes, again
increasing TC utilization.
We show how \name{} automatically derives both optimizations.

\Cref{fig:hopper-forwards-data} contains results for different
implementations of forward FMHA on Hopper.
The bars ``Triton'' and ``Triton-Tiled'' report the performance
of Triton executing the tutorial implementation of FA and
the sub-tiled version.
As discussed in \Cref{sec:exp-setup}, Triton makes many suboptimal
decisions during compilation.
The ``CUDA-Default'' bar reports the performance of our
``hand-compilation'' of Triton IR into CUDA C++, which yields
a moderate speedup over Triton.
The first \name{} result we discuss is the ``\name{}-SWP'' bar, which just uses
\name{}'s modulo scheduling (i.e., only running \textsc{Optimal-Modulo-Schedule}
in \Cref{alg:workflow}) to software pipeline the main loop and applies Triton's
WS strategy, which puts loads from global memory on a separate warp than the rest of the computation.
%
%
This results in a significant performance boost, coming within 1\% of the
official FA3 implementation at the 16384 sequence length.
If algorithms like modulo scheduling were used within systems like Triton,
this component of FA3 would not have needed human discovery.
Finally, the ``\name{}'' result
uses \name{}'s joint formulation of SWP and 
WS (i.e., running the entirety of \Cref{alg:workflow}); \name{} derived a solution in 28 seconds.
\name{}'s functional unit capacity constraints recover the ping-pong scheduling
strategy described in FA3: since only one operation may use the TC at a time,
\name{} skews groups of warps so that one issues exponentials while the other
uses the TC.
\name{} discovers the software pipeline used in FA3,
but applies it only to one warp group, determining this to be sufficient 
to saturate the functional units.
While the final implementation of this version performed slightly worse than
\name{}-SWP, the two initially performed equivalently (roughly 645 TFLOPS/s),
but an orthogonal optimization (TMA multicasting) benefited the modulo-scheduled
implementation more.
On Hopper, multiple warps are interleaved dynamically to issue work into
the TC and other functional units, naturally gaining ILP 
and lessening the burden on the quality of static instruction scheduling.
We will see next that this is no longer true on Blackwell.

\subsubsection{Blackwell}

\begin{figure}
\centering
\includegraphics[width=\linewidth]{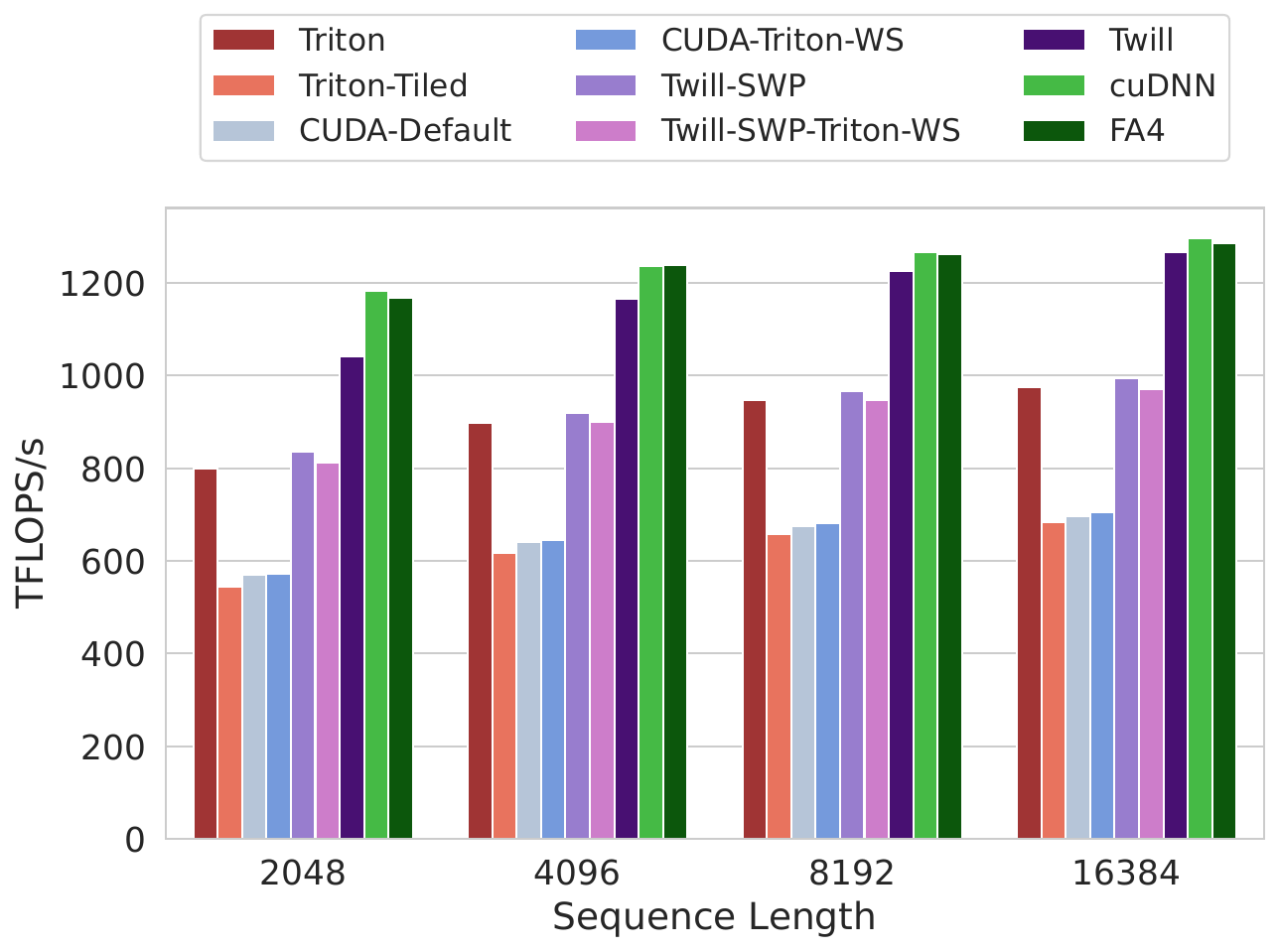}
\caption{Blackwell FP16 Non-Causal Forward Attention. \texttt{(BATCH=4, NUM\_HEADS=32, HEAD\_DIM=128)}}
\label{fig:blackwell-forwards-data}
\end{figure}

Blackwell requires substantially different SWP
and WS strategies than Hopper due to a faster TC
and a larger set of required synchronization operations (Tensor Memory
loads and stores).
The set of Blackwell results is shown in \Cref{fig:blackwell-forwards-data}.

As for Hopper, we include Triton results for the tutorial and sub-tiled
versions.
Our direct translation of Triton IR to CUDA performs worse
than Triton on Blackwell because Triton's Blackwell backend
heuristically applies the FA3 SWP strategy.
When we apply modulo scheduling and Triton's WS strategy for Hopper, 
we match Triton's performance, but are far from the best implementations.

\begin{figure}
\centering
\includegraphics[width=0.95\linewidth]{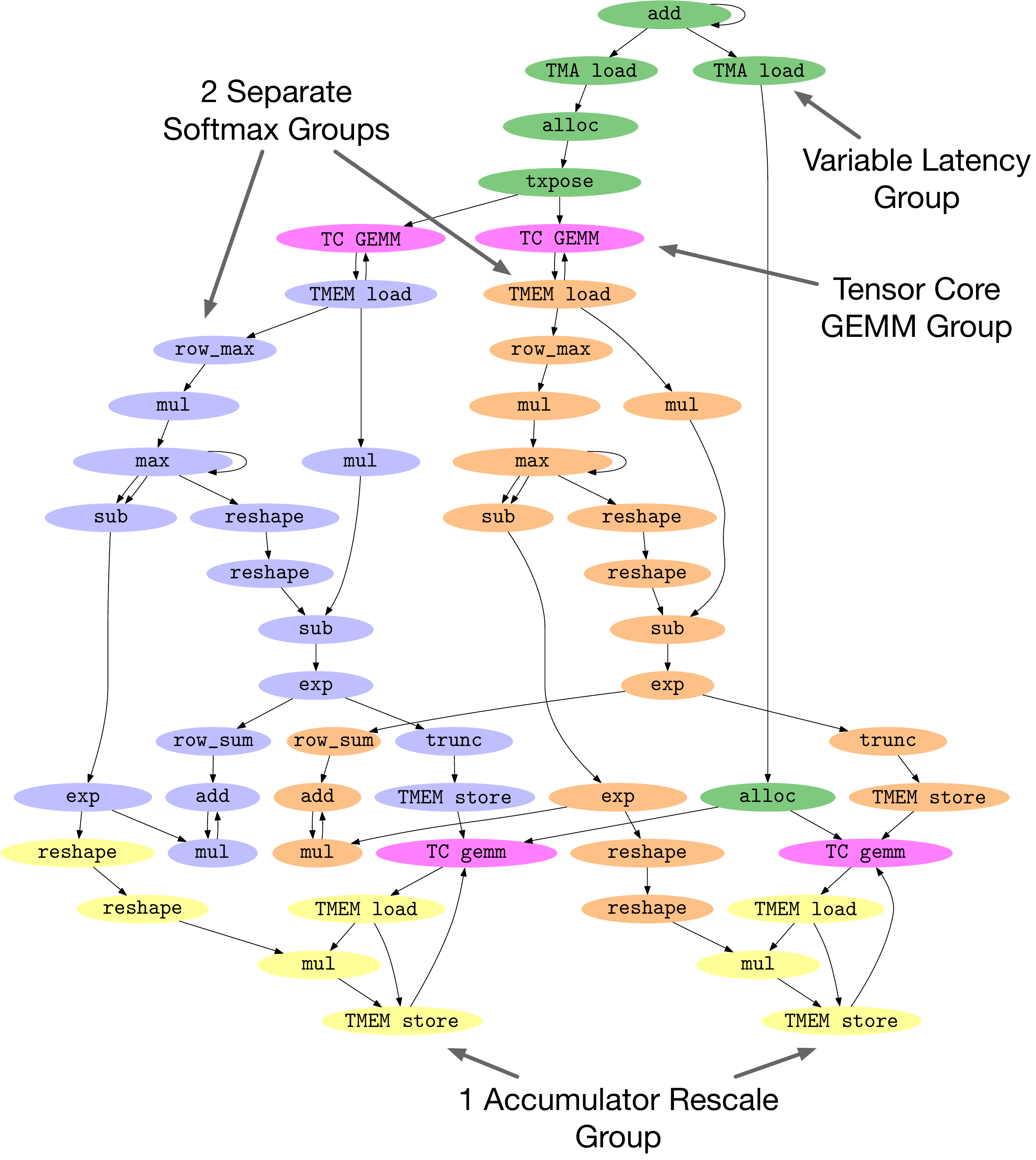}
\caption{\name{}'s dependence graph showing the discovered forward WS strategy for Blackwell. Colors indicate instructions assigned to the same group of warps.}
\label{fig:blackwell-forwards-ws-strategy}
\end{figure}

\name{}'s joint optimization of SWP and WS
completes in 19 seconds and finds a strategy that performs
significantly better than Triton and competitively with cuDNN and FA4.
In fact, the discovered strategy is exactly the same
as proposed by FA4~\cite{flash-attention-4}!
\name{}'s strategy combines a different software pipeline than Hopper
with specific warp assignments and cross-warp communication.
\name{}'s WS strategy is shown in \Cref{fig:blackwell-forwards-ws-strategy},
using a visualization generated by \name{} for the actual dependence graph
obtained from Triton IR for this program.
\name{} places variable-latency operations (green) and TC GEMMs (pink)
onto separate warps, softmax calculations for each sub-tile
onto two different groups of warps (blue and orange), and accumulator rescaling operations
for both sub-tiles onto a third group of warps (yellow).

This strategy does not comport to conventional warp roles
like ``loader'' and ``compute'' warps.
The reasoning behind this strategy is opaque without considering the associated
software pipeline.
The pipeline hoists the first GEMM for each sub-tile out of the loop like 
FA3.
In the main loop, the pipeline schedules the exponential for one
sub-tile during the GEMMs for the other sub-tile (like ping-pong scheduling),
all while the sub-tile's accumulator is being rescaled.
The rescaling is moved to a third group of warps because
reading accumulators from Tensor Memory requires blocking synchronization; 
placing it on the warp issuing exponentials disrupts the pipeline.
However, moving the rescaling to a third group of warps incurs cross-warp
communication, as it needs data produced by the softmax warps ---
\name{} schedules around this latency and synchronization.
Our implementation of this schedule comes within 2\% of FA4 on the 16384 sequence length. 

We now return to other kernel
variants in \Cref{fig:blackwell-forwards-data}.
Triton has heuristics to replicate the FA4 strategy
by moving accumulator rescaling to a separate group of warps.
Applying these heuristics to our purely software-pipelined implementation (``\name{}-SWP-Triton-WS'')
or our default translation (``CUDA-Triton-WS'') yields no benefit, showing that jointly
considering SWP and WS is critical for peak performance.

The final compelling component of our Blackwell results are modifications which
render the SMT problem unsatisfiable for the smallest $I$ which makes the \Ilp{} succeed in \Cref{alg:workflow}, forcing a search over larger values of $I$ and $L$.
These modifications include:
\begin{itemize}[nosep]
    \item Reducing the number of warps.
    \item Using the tutorial Triton implementation directly. Without sub-tiling,
    the instruction granularity is not fine enough for interleaving.
    \item Prohibiting cross-warp communication.
\end{itemize}
The variety of unsatisfiable configurations demonstrates the complexity
of the search space for finding high performance SWP and WS strategies.
\name{} provides a technical framework for reasoning about and navigating this
complex search space while guaranteeing optimal results.

\subsection{Attention Backward Pass}

We now move to the backward pass of FMHA.
%
The backward pass is significantly different from the forward pass, requiring different
SWP and WS strategies.
%
%
We did not use the Triton tutorial implementation of the
backward pass, as it uses a work-inefficient two-pass algorithm.
Instead, we implemented the single-pass backward algorithm as described in
FA3~\cite{flash-attention-3} in Triton, allowing us to directly compare against 
cuDNN and the reference FA implementations.
The single-pass algorithm consists of five GEMMs and one exponential, followed by
an atomic reduction into global memory.
Similar to the forward pass, we also implement a sub-tiled version of the algorithm.

\subsubsection{Hopper}

\begin{figure}
\centering
\includegraphics[width=0.90\linewidth]{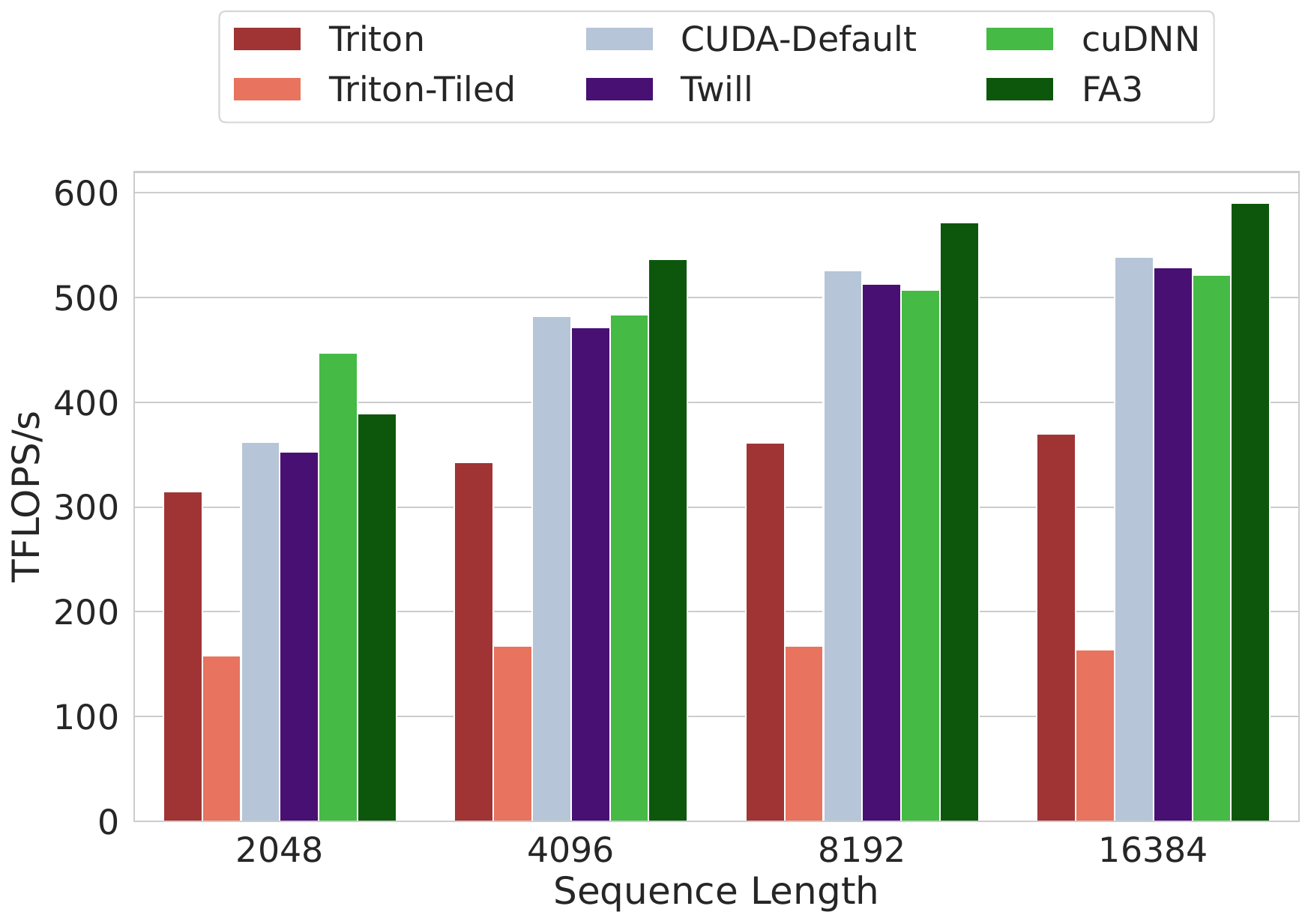}
\caption{Hopper FP16 Non-Causal Backward Attention Results. \texttt{(BATCH=4, NUM\_HEADS=32, HEAD\_DIM=128)}}
\label{fig:hopper-bwd-results}
\end{figure}

Hopper results are shown in \Cref{fig:hopper-bwd-results}.
Similar to forward on Hopper, our port of Triton IR
into CUDA achieves a performance boost over Triton and performs
competitively with cuDNN on the highest sequence length.
Backward attention on Hopper is register constrained, and thus
FA3~\cite{flash-attention-3} does not construct a software pipeline that
exploits ILP across iteration boundaries.
\name{} searched for 88 seconds and found a similar schedule: it exploited ILP
within a loop iteration but could not do so across iterations due to register
capacity.
%
\name{}'s findings confirm that the developers
of FA3 did not miss potential performance with their software pipelining strategy.
%
We believe the reference FA3 implementation beats ours
by 11\% on the 16384 sequence length by using a slightly larger 
tile size (80x128), allowing for the use of larger GEMMs.
Triton only supports power-of-two tile sizes, so we were not able to construct
IR with the same tile size, but there is no fundamental reason that 
\name{} cannot handle tile sizes that are not powers of two.

\subsubsection{Blackwell}

\begin{figure}
\centering
\includegraphics[width=0.9\linewidth]{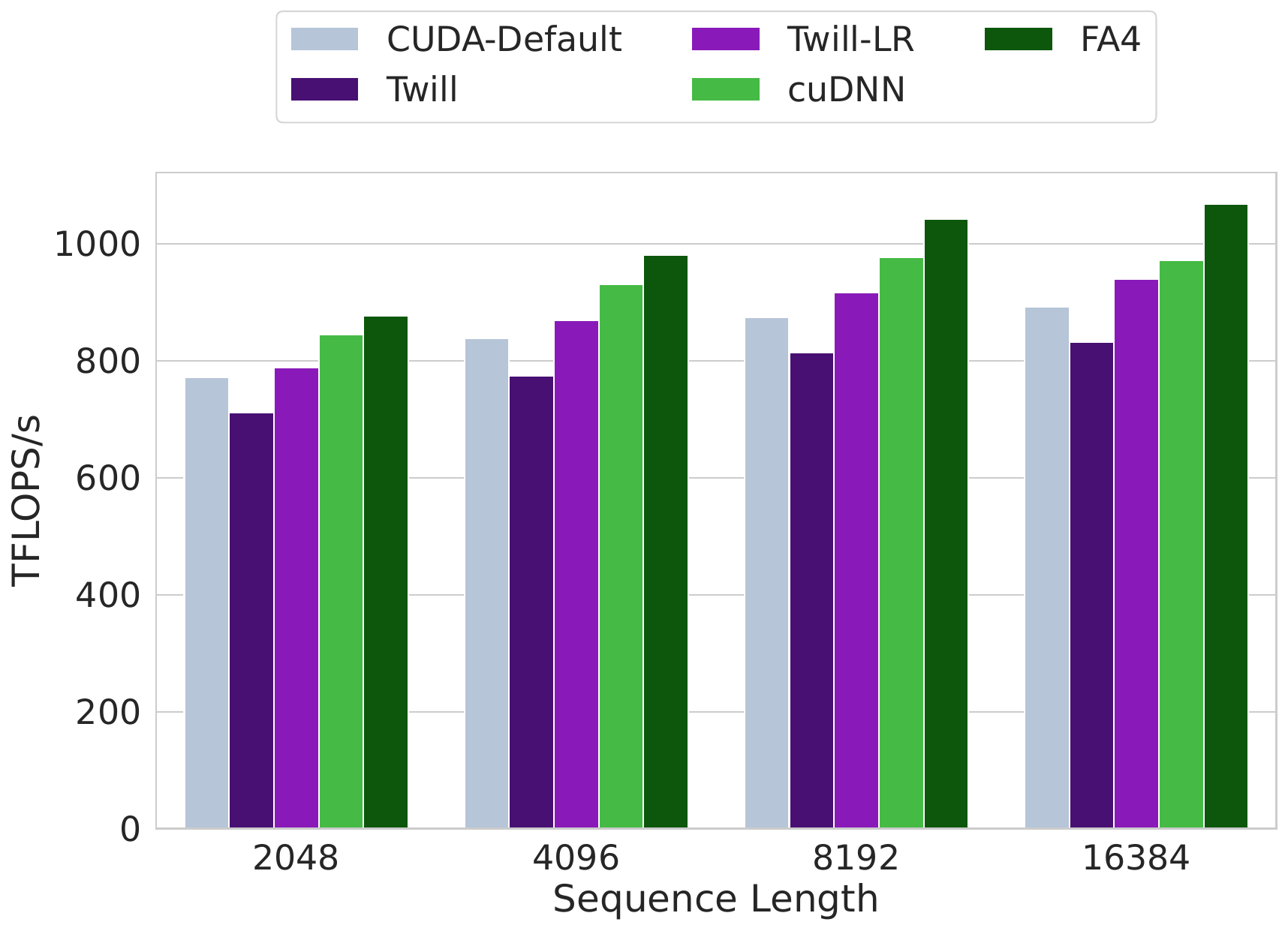}
\caption{Blackwell FP16 Non-Causal Backward Attention Results. \texttt{(BATCH=4, NUM\_HEADS=32, HEAD\_DIM=128)}}
\label{fig:blackwell-bwd-results}
\end{figure}

Blackwell results are shown in \Cref{fig:blackwell-bwd-results}.
Triton was unable to generate code for Blackwell, as it is currently unable
to construct Tensor Memory allocation strategies that contain aliasing.
%
%
The FA4 strategy for Blackwell~\cite{flash-attention-4} uses
three groups of warps:
two groups read accumulators from Tensor Memory to apply exponentials, and the last group
stages accumulators in shared memory for the outgoing atomic reduction.
\name{} finds a different strategy in 269 seconds using only two 
groups of warps which ping-pong between performing exponentials and staging 
reduction data (``\name{}'').
Despite our own inspection confirming the schedule should fit within
the register-per-thread budget, {\tt ptxas} was unable to register allocate this code
without significant register spilling in the main loop, degrading performance.
We then ran \name{} with a reduced register-per-thread budget, and
in 64 seconds it found the same strategy as described earlier that 
used three groups of warps.
{\tt ptxas} succeeded register allocating this version, yielding the small speedup over
the default implementation (``\name{}-LR'').
The overall difference between Blackwell backward implementations is relatively small;
the high throughput of the Blackwell TC renders the single-pass algorithm
bandwidth-bound, limited by the completion of the atomic reduction at each iteration.
However, \name{} finds a similar SWP and WS strategy that
hides some exposed latency present in a direct port of the Hopper backward strategy.
We believe the remaining performance differences to be from orthogonal optimizations around
memory layouts and instruction selection in the well-tuned reference implementations.

\section{Related Work}

\subsection{Tensor Core GPU Programming}

The combination of widely varying expertise in GPU programming
and the desire for diverse and high-performance machine learning
computations has led to a large ecosystem of programming systems
for targeting TC GPUs.
We provide a survey of this space of systems.

Below deep-learning frameworks like PyTorch~\cite{pytorch} or JAX~\cite{jax},
domain-specific languages (DSLs) like Triton~\cite{triton}, Pallas~\cite{pallas}
and CUDA Tile~\cite{cutile} provide abstractions for operating on tiles of data.
The compilers for each language are then responsible for generating
efficient code, often achieving moderate to high performance.
Slightly lower-level DSL's like TLX~\cite{tlx} and TileLang~\cite{tile-lang}
also expose tile-based programming models, but provide more control
over optimizations.
Gluon~\cite{gluon} is a Triton extension that exposes the Triton low-level 
intermediate representation (IR) to the programmer.
Gluon programmers gain some automation from further lowering of this
IR, but retain control over optimizations
like memory allocation and WS.
Cypress~\cite{cypress} is a task-based programming model that
allows programmers to express the decomposition of computation
and data across the GPU.
Cypress provides a separate mapping interface to
control performance-sensitive decisions without 
affecting correctness.
Finally, systems like ThunderKittens~\cite{thunderkittens} and
CUTLASS~\cite{cutlass} provide the developer complete control,
leaving no performance-sensitive decisions to an automated system.

\subsection{Software Pipelining}

\paragraph{Pipelining for CPUs.}
Software pipelining is a classic and well-understood optimization
for CPUs, critical
for extracting enough instruction level parallelism to saturate the machine.
Even for modern out-of-order processors, software pipelining can
improve instruction throughput by filling the processor's reorder
buffer with independent instructions.
Compaction-based~\cite{aiken-pipelining} pipelining algorithms
logically unrolled the target loop until a fixed-point pipeline is
discovered.
Modulo scheduling~\cite{lam-modulo-scheduling, rau-modulo-scheduling} pipelining
algorithms model steady state resource usage with a modulo table data structure.
Modulo scheduling algorithms are amenable to the derivation of optimal solutions
through the use of Integer Linear Programming~\cite{gao-ilp-scheduling, stoutchinin-ilp}.
\name{} shows how to adapt traditional modulo scheduling techniques for Tensor
Core GPUs.

\paragraph{Pipelining for Tensor Core GPUs.}
ALCOP \cite{alcop} does SWP of loads with compute, and is therefore insufficient
to derive pipelines that overlap compute operations, like FA3.
PipeThreader \cite{pipethreader} finds an SWP strategy by defining a space
of programs and then searching over it, profiling each candidate to find the
fastest.
In contrast, \name{} takes a constraint-based approach where the space is
implicitly defined by the constraints, and the search is executed by
highly-optimized solvers.
Moreover, \name{} does not rely on profiling---our approach is static and
instead relies on publicly available specifications to estimate operation
latencies.
Furthermore, PipeThreader relies on the conventional, heuristic approach to WS
involving the creation of producer and consumer warps, whereas \name{} derives a
WS strategy from first principles.
Most importantly, PipeThreader does not provide any optimality guarantees
regarding the schedules it finds.

\subsection{Warp Specialization}

The GPU programming technique of warp specialization (WS) was introduced
before NVIDIA GPUs contained TCs.
CUDA-DMA~\cite{cuda-dma} leveraged WS to separate data movement
between global and shared memory from computation, achieving higher memory bandwidth.
Singe~\cite{singe} partitioned combustion chemistry computations across
warps to shrink the register-level working set of each warp, allowing large
chemical reactions to be simulated without spilling.

Since Hopper, WS has become ubiquitous in high-performance TC kernels.
Programming systems for TC GPUs offer many different layers of control
around WS.
High-level systems like Triton~\cite{triton}, Cypress~\cite{cypress}
or Tawa~\cite{tawa} heuristically define a WS
\emph{strategy}, and then automatically transform the 
source program to implement chosen strategy.
Mid-level systems like TLX~\cite{tlx} allow specifying
a WS strategy through source-level annotations,
and then implement the strategy through transformation
algorithms within the compiler.
Finally, low-level systems like Gluon~\cite{gluon}, ThunderKittens~\cite{thunderkittens}
or CUDA C++ require users to both define a WS
strategy and realize it in low-level code.
\name{} focuses specifically on the problem of defining a WS
strategy, which all other existing systems perform
through machine-specific heuristics or by relying on programmer intuition.

\section{Conclusion}

We have presented \name{}, a system that discovers optimal
SWP and WS strategies for Tensor Core GPUs.
\name{} presents a novel joint formulation of SWP and
WS that can be offloaded to  \Ilp{} and SMT solvers, deriving the complex
strategies that experts have  found by 
hand for multiple generations of GPUs.
We additionally show that approaches that consider these optimization
strategies separately are unable to achieve peak performance.
%

\section*{Acknowledgments}
We thank Duane Merrill for discussions about GPU architecture
and inspiring our view of warp specialization.
We thank Vinod Grover for discussions about warp specialization and
software pipelining.
We thank Evghenii Gaburov, Masahiro Masuda and Jason Knight for their support
with Triton development and Triton's warp specialization infrastructure.
We thank Pradeep Ramani and Cameron Shinn for their help with performance
debugging Hopper and Blackwell kernels.
We thank Benjamin Driscoll for suggesting the acronym \Ilp{}.
We thank (in no particular order) Chris Gyurgyik, Katherine Mohr, AJ Root, 
Shiv Sundram, Atharva Chougule, Sai Gautham Ravipati and Konstantinos Kallas for providing 
feedback on drafts of this manuscript.




\bibliographystyle{plain}
\bibliography{main}

\end{document}